\documentclass[usenatbib]{mn2e}
\usepackage[dvips]{graphicx}

\title[CMB and foregrounds in WMAP first year data]
{CMB and foregrounds in WMAP first year data.}
\author[G. Patanchon, J.-F. Cardoso, J. Delabrouille, P. Vielva]
{G. Patanchon$^1$, J.-F. Cardoso$^2$, J. Delabrouille$^3$, P. Vielva$^3$ \\
$^1$ Department of Physics \& Astronomy, University of British Columbia,
6224 Agricultural Road, Vancouver, BC V6T 1Z1, Canada\\
$^2$ CNRS/ENST --- 46, rue Barrault, 75634 Paris, France\\
$^3$ PCC --- Coll{\`e}ge de France, 11, place Marcelin Berthelot,
F-75231 Paris, France\\
}
\date{Accepted ??. Received ??; in original form ??}
\pagerange{\pageref{firstpage}--\pageref{lastpage}}
\pubyear{200x}
\begin{document}

%---------------

\maketitle
\label{firstpage}

\begin{abstract}
  We perform a blind multi-component analysis of the WMAP 1 year
  foreground cleaned maps using SMICA (Spectral Matching Independent
  Component Analysis). We provide a new estimate of the CMB power
  spectrum as well as the amplitude of the CMB anisotropies across
  frequency channels. We show that the CMB anisotropies are compatible
  with temperature fluctuations as expected from the standard
  paradigm. The analysis also allows us to identify and separate a
  weak residual galactic emission present significantly in the Q-band
  outside of the Kp2 mask limits, and mainly concentrated at low
  galactic latitudes.  We produce a map of this residual component by
  Wiener filtering using estimated parameters. The level of
  contamination of CMB data by this component is compatible with the
  WMAP team estimation of foreground residual contamination. In
  addition, the multi-component analysis allows us to estimate jointly
  the power spectrum of unresolved point source emission.
\end{abstract}

\begin{keywords}
  Cosmic microwave background -- Cosmology: observations -- Methods:
  data analysis
\end{keywords}

%------------------------------------------------------

\def\bR{\mathbf{R}}
\def\bA{\mathbf{A}}
\def\bs{\mathbf{s}}
\def\bD{\mathbf{D}}
\def\bn{\mathbf{n}}
\def\bN{\mathbf{N}}
\def\bx{\mathbf{x}}
\def\bW{\mathbf{W}}
\def\bC{\mathbf{C}}

\def\diag{\mathrm{diag}}
\def\trace{\mathrm{trace}}
\def\inv{^{-1}}
\def\adj{^\dagger}
\def\degree{^\circ}

\def\lmin{\ell_{\rm min}}
\def\lmax{\ell_{\rm max}}

%------------------------------------------------------

\section{Introduction}

The Cosmic Microwave Background (CMB) is one of the most powerful
probe of modern cosmology. The shape of the spatial power spectrum of
the small temperature fluctuations depend on the cosmological
parameters describing, in the frame of the standard model, the matter
content, the geometry, and the evolution of the Universe
\citep{jungman96}. Since the first detection of CMB anisotropies by
the COBE satellite in 1992 \citep{DMR}, several ground--based and
balloon--borne experiments have provided an accurate estimate of the
power spectrum on a large range of angular scales.  The recent WMAP
mission \citep{WMAP1yr}, after one year of data acquisition, provided
measurements of the power spectrum with unprecedented accuracy.

Measuring the CMB power spectrum is a difficult task. It requires a
good characterization of noise contribution in the observations, as
well as the subtraction of foreground astrophysical emission present
at millimeter wavelengths \citep{fb-rg99}.  Both noise and foreground
emissions can significantly bias CMB power spectrum estimates if not
well accounted for.

The WMAP data is the best presently available observation of the sky
in the 20-90~GHz range.  A careful separation of CMB from foregrounds,
however, is required in order to make the best out of these
observations.  In particular, isolating CMB from foregrounds is of
extreme importance for measuring accurately the angular power spectrum
$C(\ell)$ of CMB anisotropies and for cosmological interpretation.

The approach of the WMAP collaboration consists in cleaning
observations from foreground contaminations for performing a CMB power
spectrum estimation on the cleaned maps. More specifically, galactic
foregrounds are subtracted using templates obtained from external data
sets.  In addition, strong known extra-galactic point sources, as well
as the region of the galactic plane, are masked prior to the analysis.
The CMB spectrum is estimated using a weighted average of the
cross-power spectra of cleaned observation maps in the Q, V and W
channels \citep{WMAPSp}.  This avoids biases due to detector noise,
assuming that the noise of different detectors is uncorrelated.  This
estimated spectrum is corrected for residual point source
contamination by subtracting an estimate of the contribution of point
sources in the cross power spectra. The level of residual galactic
contaminations is estimated by cross-correlating the maps with
foreground templates.

A multi-component approach of CMB spectral estimation has been
proposed by \cite{SM-mnras}. Additional publications on the subject
give variants and details \citep{cardoso02,patanchon03}.  The method,
called SMICA (Spectral Matching Independent Component Analysis), is
based on matching the cross- and auto-power spectra of the observed
maps to a parametric model described by the power spectra of all the
astrophysical components, their relative amplitudes in the different
channels, and the noise power spectra.
It is a very flexible approach: depending on available knowledge, most
of the parameters may be either estimated from the data or kept at
fixed values.  In particular, emission laws for all or some of the
components can be estimated.

There are several benefits to using SMICA in CMB analysis.

First, SMICA is equivalent to a maximum likelihood estimation of the
CMB power spectrum if the observed maps are a linear mixture of
Gaussian stationary components and noise. Therefore, if there are no
foregrounds but only Gaussian CMB and noise, SMICA is expected to
outperform quadratic estimates.

Secondly, SMICA allows to estimate jointly the spatial power spectra
\emph{and} the amplitude of components in each channel (which are
related to emission laws of components as well as calibration
coefficients).
In particular, the application on WMAP data allows us to check the
following strong prediction of the standard model: CMB anisotropy
should have a spectral emission law given, to first order, by a
derivative (with respect to the temperature) of the blackbody law.
\cite{fixsen03} has measured the CMB anisotropy contribution in the
COBE FIRAS data by looking for the WMAP anisotropy template in FIRAS.
The author shows the compatibility of CMB anisotropy at large angular
scales ($> 5^\circ$) with temperature fluctuations in the Wien part of
the spectrum. Our blind estimation of the component emission laws
provides us with a unique tool for extending this result to
frequencies covered by WMAP and to all measured angular scales
(including the first acoustic peak) with a remarkable precision as
shown below.

Finally, if there are foreground contributions in the observations,
SMICA is designed to detect them and allows their separation.  It
permits indeed to jointly estimate power spectra of multiple
components in the data, {\emph{to assess the significance of
components}}, and eventually to separate the effect of all the
emissions.

\medskip

In this paper, we investigate the existence of residual foreground
emission in foreground--cleaned WMAP data.  In particular, residual
galactic emission resulting from an error in the subtraction of
galactic templates can be present in the published maps. Such
residuals may exist if external templates of galactic emissions,
extrapolated using a physical model to WMAP frequencies, are not quite
representative of the actual galactic components. Multi-component
analysis with SMICA allows to check for the existence of such a
residual, and to quantify the level of contamination if any, without
prior information on its power spectrum or on its amplitude across
channels.

In addition, the WMAP collaboration has shown that unresolved point
source emissions have a non-negligible contribution to the data
\citep{WMAPSp,komatsu}. This result has been independently confirmed
by \cite{SZ-WMAP}. As will be shown, SMICA also yields a coarse
estimate the level of point source emission in WMAP data.

The paper is organized as follow: section \ref{sec:data} introduces
the data used for the analysis, the simulations performed in order to
check our results and the model of the WMAP observations for SMICA.
The method is described in section \ref{sec:method}. Section
\ref{sec:results} presents the different results. Finally,
conclusions are provided in section \ref{sec:conclusion}

\section{Data}\label{sec:data}

\subsection{WMAP data and input maps for SMICA}\label{sub:maps}

The WMAP space probe, launched by NASA in 2001, is a large telescope
for imaging the total emission of the sky at 5~different wavelengths
(or frequency channels), with a resolution ranging from about 0.2
degrees to 0.9 degrees (limited by diffraction), and with full sky
coverage \citep{WMAP1yr}.

The data taken by WMAP has been made available to the scientific
community after one year of proprietary period, and is freely
available on a dedicated NASA CMB web
site.\footnote{http://lambda.gsfc.nasa.gov/} The data consists in a
set of 10 maps obtained by different detector pairs: four maps at 3.2
mm (W band at 94 GHz), two at each 4.9 mm and 7.3 mm (V band at 61 GHz
and Q band at 41 GHz), and one at 9.1 mm and at 13.0 mm (Ka band at 33
GHz and K band at 23 GHz). The data is provided in the HEALPix
pixellisation format of the
sphere\footnote{http://www.eso.org/science/healpix/} \citep{healpix}.
Foreground cleaned versions of the Q, V and W-band maps are also
available. These maps have been used in the generation of the WMAP
first-year CMB power spectrum by the WMAP team. The Galactic
foreground signal, consisting of synchrotron, free-free, and dust
emission, was removed using the 3-band, 5-parameter template fitting
method described in \cite{wmap-foreg}. Galactic templates come from
external data sets: the 408 MHz synchrotron map \citep{haslam82}, the
predicted dust emission at 94 GHz using the FDS model
\citep{finkbeiner99}, composite H$\alpha$ map \citep{finkbeiner03} and
the galactic reddening E(B-V) \citep{schlegel98}.

For our analysis, we use the eight individual foreground cleaned maps.
We partially correct the maps from beam smoothing effects, so that
every map gets the effective resolution of a reference map (we choose
W3). We simply multiply the coefficients of the spherical harmonic
decomposition by the inverse of beam transfer function ratio.

We consider two different sets of maps corresponding to different
masking. For the first set (hereafter map set I), we apply the Kp2
galactic mask provided by the WMAP collaboration, canceling about 15\%
of the pixels. For the second set (hereafter map set II), we mask the
pixels corresponding to galactic latitudes lower than 40 degrees. For
both sets, we mask the strongest point sources using the mask provided
by the WMAP team. In all cases masks are apodised for a smooth
transition between 0 and 1 on a scale of 30 arcmin.

The prior correction of beam smoothing effects permits the application
of SMICA using directly the pseudo cross- and auto-power spectra of
observation (computed from partially covered maps).  The true component power  
spectra
can be obtained from these pseudo power spectra afterwards.

\subsection{Simulations}

In order to validate the method, assess error bars on the estimated
parameters, and check for systematic errors in the analysis, we have
generated 100 full-sky simulations of the WMAP observed maps with 6.5
arcmin pixels (HEALPix nside = 512). Each set of simulations consists
in 8 maps reproducing the observations in the Q, V and W bands. Maps
contain synthetic CMB anisotropies degraded to the resolution of the
detectors, assuming a symmetric beam pattern and using the transfer
function published by WMAP.  They contain anisotropic white noise at
the expected level.  CMB anisotropies are generated using the CMBFAST
software \citep{2000ApJS..129..431Z} with current `concordance'
cosmological parameters \citep{spergel03}. The simulated maps are
partially deconvolved and the same mask as used for real data is
applied.

No measurable bias has been observed in simulations.  Error bars on
all parameters (CMB power spectrum as well as mixing parameters)
obtained from the dispersion over 100 independent simulations have
been found to be in very good agreement (precision better than 10\%)
with the closed form expression of eq.~(\ref{eq:crb}), validating the
use of the latter for final error bar estimates.

\subsection{Model of the WMAP observations}\label{subsub:model}

The sky emission at WMAP frequencies is well described to first order
by a linear superposition of the emissions of a few processes: CMB
anisotropies, galactic foregrounds (synchrotron, dust,
free-free\ldots), point source emissions and Sunyaev Zel'dovich (SZ)
effect.  After subtraction of the galactic foreground templates and
masking, the WMAP data between 40 and 94 GHz can be modeled as noisy
linear mixtures of CMB anisotropies, unresolved point sources and
possibly small residual galactic emissions (no significant SZ effect
is expected to be present in WMAP observations \citep{wmap-foreg}, and
\cite{SZ-WMAP} found no evidence of SZ effect using a multi-frequency
approach).  We assume that the emission laws of components are
independent of the position ($\theta,\phi$) on the sky.
Although this is only an approximation, it holds exactly for the CMB
and for galactic residuals which are significant at one frequency
only.  Thus, observation $i$ is modeled as
\begin{equation}
x_i(\theta,\phi)=\sum_{c=1}^{N_c}A_{ic}{s}_c(\theta,\phi)+n_i(\theta,\phi)
  \label{eq:mixture}
\end{equation}
where ${s}_c$ is the spatial distribution of component $c$, $n_i$ is
the noise of observation $i$ and $A_{ic}$ is the amplitude of
component $c$ in map $i$ given by 
\begin{equation}
  A_{ic} = \int w_i(\nu)~ g_c(\nu) d\nu \label{eq:mixmat}
\end{equation}
where $w_i$ is the spectral band of detector $i$ and $g_c(\nu)$ is the
emission law of component $c$.
In matrix-vector format, model~(\ref{eq:mixture}) reads
\begin{equation}
  \label{eq:mixturemv}
  x(\theta,\phi)
  =
  A s (\theta,\phi) + n(\theta,\phi)
\end{equation}
with an $N_d\times N_c$ mixing matrix $A$.

This simple model does not reflect the fact that resolution depends on
the frequency band.  In our analysis, this effect, as well as the
impact of the masks, is accounted for in the spectral domain: see
section~(\ref{sec:beam-cover-effects}), for a sketch of the correction
of beam and coverage effects.

On the statistical side, we shall assume statistical independence
between components, between the noise contaminations of different
detectors, and between the latter and all the components.

The standard model of cosmology predicts that CMB anisotropies are
small temperature fluctuations of a pure blackbody spectrum. A first
order expansion around $T=2.726\,{\rm K}$ gives:
\begin{equation}
  g_{_{\rm CMB}}(\nu)
  \propto
  \left[\frac{\partial B_\nu(T)}{\partial T}\right]_{T=2.726\,{\rm K}}
\end{equation}
where $B_\nu(T)$ is the Planck law for the emission of a blackbody at
temperature $T$.

Whereas the assumption that $A_{ic}$ does not depend on the position
is a good approximation to first order for galactic components, it is
not the case for point source emissions. Nevertheless, since the
brightest point sources are masked and since the frequency dependence
of most of contributing radio sources belong to the flat population,
their emission law can be roughly described as:
\begin{equation}
  g_{_{\rm PS}}(\nu) \propto \left({\nu \over \nu_0}\right)^{\beta}
\end{equation}
with $\beta \approx -2$ \citep{toffolatti98}.
Note that unresolved point sources in the WMAP data have already been
reported at a very low level~\citep{WMAPSp}.

\subsection{Spectral statistics}

The spectra and cross-spectra at frequency $\ell$ of $x(\theta,\phi)$
(considered as an isotropic $N_d$-dimensional random field on the
sphere) is the $N_d\times N_d$ spectral matrix $R(\ell)$
\begin{equation}
  R(\ell) = \langle\, x(\ell,m)\, x(\ell,m){^\dagger}\ \rangle
\end{equation}
where $\langle\cdot \rangle$ denotes an ensemble average, superscript
$^\dagger$ denotes transposition and the $x(l,m)$ are the coefficients of the
field on the basis of real spherical harmonics.
This average value is independent of $m$ because of isotropy.

In the linear model of eq.~(\ref{eq:mixturemv}), independence between
components and noise implies that
\begin{equation}
 R(\ell) = A C(\ell) A^t + N(\ell) \label{eq:sptheo}
\end{equation}
where $C(\ell)$ and $N(\ell)$ are the spectral matrices for the
components and the noise respectively.
They are both diagonal matrices as a consequence of the independence
assumption between components and between noise contaminations.

In practice, spectral matrices are estimated by averaging over
frequency bins.
Typically, one considers $Q$ frequency bins with the $q$-th bin
($1\leq q\leq Q)$ containing all harmonic modes $(\ell,m)$ such that
$\lmin (q) \leq \ell \leq \lmax (q)$.
If $n_q= \sum_{\ell=\lmin (q)}^{\lmax (q)} (2\ell+1)$ denotes the
number of modes in the $q$th harmonic bin, then the empirical spectral
matrix
\begin{equation}\label{eq:binexp}
  \widehat R_q
  =
  \frac 1 n_q
  \sum_{\ell=\lmin (q)}^{\lmax (q)}\sum_{m=-\ell}^{\ell}  
x(\ell,m)x(\ell,m)^\dagger
\end{equation}
is the natural estimate of the average spectral matrix
\begin{equation}\label{eq:avspecmat}
  R_q
  =
  {1 \over n_q}
  \sum_{\ell=\lmin (q)}^{\lmax (q)}  (2\ell+1) R(\ell)
  .
\end{equation}
The latter inherits its structure from (\ref{eq:sptheo}) as:
\begin{equation}\label{eq:bintheo}
  R_q= A C_q A^t + N_q
\end{equation}
with average spectral matrices $C_q$ and $N_q$ related to $C(\ell)$
and $N(\ell)$ as in definition~(\ref{eq:avspecmat}).

\section{Spectral estimation and component separation with SMICA}  
\label{sec:method}

The goal of our analysis is to identify and separate the components
present in WMAP maps, to evaluate the amplitude of CMB anisotropies as
a function of observation wavelength (as given by the corresponding
column of $A$), and to provide accurate estimates for power spectra of
the CMB and other components.
This section briefly describes a multi-detector, multi-component
analysis method.  It is a maximum likelihood method based on a model
of statistically independent components.  Since it can be understood
as a spectral matching technique, it is dubbed ``SMICA'', standing for
``spectral matching independent component analysis''.
More details on SMICA can be found in \cite{SM-mnras}.

\subsection{Spectral matching}\label{sub:method}

The SMICA analysis technique consists in minimizing a measure of the
mismatch between the empirical covariance matrices $\widehat R_q$
and their model counterparts $R_q$.  Minimization is conducted with
respect to any relevant set of parameters describing the covariance
matrices $R_q$.
The maximal parameter set is made of: the entries of the mixing matrix
$A$, the average power $[C_q]_{cc}$ of the $c$-th component in the
$q$-th bin, and the average power $[N_q]_{ii}$ of the noise on the
$i$-th detector in the $q$-th frequency bin.

One may choose to fix some of these quantities (prior knowledge) and
to estimate the remaining ones, either freely or under some additional
parametric constraints (see section \ref{sub:choiceparam}).  In any
case, denoting by $\theta$ a set of free parameters defining the
values of $A_{ic}$, $[C_q]_{cc}$ and $[N_q]_{ii}$, we obtain an
estimate of $\theta$ by minimizing a joint spectral mismatch,
quantified by a weighted sum
\begin{equation}
  \label{eq:phi}
  \Phi(\theta) = \sum_{q=1}^Q n_q  D
  \left(
    \widehat R_q, R_q(\theta)
  \right)
  .
\end{equation}
Any sensible measure $D(\cdot,\cdot)$ of matrix mismatch could be used
to obtain a consistent estimate $\hat\theta$ of the $\theta$ parameter
as $\hat\theta=\arg\min_\theta \Phi(\theta)$.  We adopt
\begin{equation}
  \label{Kullback}
  D(R_1,R_2) = \frac12 \left[
    \trace \left( R_1R_2\inv \right) - \log\det (R_1R_2\inv) - N_d
    \right]
\end{equation}
because then criterion~(\ref{eq:phi}) is, up to a constant, equal to
minus the log-likelihood of the data in a simple Gaussian isotropic
model (see \cite{SM-mnras} for a derivation).  Hence, minimizing
criterion~(\ref{eq:phi}) is equivalent to maximizing the likelihood.
This fact guarantees good performance (at least for large $n_q$) when
the data do come from an isotropic random field.  Also, the connection
with the likelihood criterion suggests a simple optimization strategy
using the EM (expectation-maximization) algorithm.
In practice, minimization of the spectral mismatch~(\ref{eq:phi}) is
achieved with the EM algorithm followed by few steps of a descent
method (BFGS algorithm) to speed-up the final convergence.

\subsection{Error estimation}

The SMICA estimator being a maximum likelihood estimator, the
asymptotic variance-covariance matrix of the estimates is given, when
the model holds for some value $\theta_0$ of the parameters, by 
\begin{equation}\label{eq:crb}
  \mathrm{Cov}(\hat \theta)
  =
  \langle 
  \{(\hat\theta-\theta_0)(\hat\theta-\theta_0)^t\}
  \rangle
  \approx
  J(\theta_0) ^{-1}
\end{equation}
where the Fisher information matrix $J(\theta_0)$ is, in our model
\begin{equation}
  [ J(\theta_0) ]_{ij}
  =
  \frac12 \sum_q  n_q\, \trace
  \left\{
    R^{-1}_q
    {\partial R_q\over \partial\theta_i}
    R^{-1}_q
    {\partial R_q\over \partial\theta_j}
  \right\}
\end{equation}
with all quantities evaluated at point $\theta_0$.  In practice,
$J(\theta_0)$ is approximated by replacing, for each frequency bin
$q$, matrix $R_q$ by $\widehat R_q$ and matrix ${\partial R_q\over
\partial\theta_i} (\theta_0)$ by ${\partial R_q\over \partial\theta_i}
(\hat\theta)$.

Departure from stationarity\footnote{In particular, as a consequence
of WMAP scanning strategy, the noise variance per pixel on the
observed maps is higher by a factor of the order of 4 around the
ecliptic equator as compared to the pole regions.}  and Gaussianity
does not introduce bias in parameter estimation but is likely to
induce a larger estimation error than predicted by the above formula.
The accuracy of the error bars prediction can be checked thanks to
simulations.

The galactic and point source mask has the effect, to first order, of
decreasing the effective number $n_q$ of modes in each bin by a factor
equal to the fraction of sky coverage.  This is accounted for by
multiplying every parameter error estimate by the inverse square root
of this factor.  This procedure has been validated with the help of
numerous Monte-Carlo simulations.

\subsection{Mismatch measure}\label{sub:Kfit}

The SMICA approach includes a built-in goodness of fit: the departure
of the spectral statistics $\widehat R_q$ from the best fit model of
independent components is quantified by the divergence $D ( \widehat
R_q, R_q(\hat\theta) )$ which should be statistically small.
For a known $N_d\times N_c$ mixing matrix, if the model of
observations is correct, then $2 n_q D (\widehat R_q,
R_q(\hat\theta))$ is asymptotically (\textit{i.e.}  for $n_q$ large
enough) distributed as a $\chi^2$ with $N_d(N_d+1)/2-(N_c+N_d)$
degrees of freedom.  This value is obtained by subtracting the number
of adjustable parameters per frequency bin ($N_c$ auto-spectra plus
$N_d$ noise levels) from the $N_d(N_d+1)/2$ degrees of freedom of a
symmetric matrix.  In particular:
\begin{equation}
  \label{eq:Kfit}
  \langle 
  n_q D
  (
  \widehat R_q, R_q (\hat\theta)
  )
  \rangle
  =
  \frac12 \left[ \frac{N_d(N_d+1)}{2} - (N_c+N_d) \right].
\end{equation}

\subsection{Beam and coverage effects}\label{sec:beam-cover-effects}

We discuss the corrections needed to account for beam and coverage
effects.

In first approximation, if a mask covers a fraction $\alpha$ of the
sky, an effective number $\alpha n_q$ should be used in place of
$n_q$.  However, since masking also introduces mode coupling, a better
approximation is desired.  We follow the \textsc{MASTER} formalism
developed by \cite{master}.  The idea is as follows: let
$s(\theta,\phi)$ be an isotropic Gaussian random field with harmonic
spectrum $c(\ell)$, of which only a masked version $\tilde
s(\theta,\phi) = w(\theta,\phi) s(\theta,\phi)$ is observed.  Denoting
$\tilde s(\ell,m)$ the harmonic coefficients of $\tilde
s(\theta,\phi)$, we have
\begin{equation}
  \label{eq:MASTER}
  \frac1{2\ell+1}
  \sum_{m=-\ell}^{m=\ell}
  \langle | \tilde s(\ell,m)|^2 \rangle
  =
  \sum_{l'} M_{\ell\ell'}  c(\ell')
\end{equation}
where the coefficients $M_{\ell\ell'}$ depend only on the mask (and
thus can be computed independently of the data).
Hence, if $M_{\ell\ell'}$ is known, so is the bias introduced by the
mask on the harmonic spectra.

In practice, we proceed as follows.  In a first step, the input maps
are masked, transformed to the spherical harmonic domain and brought
to a common resolution (see section \ref{sub:maps}).  Next,
bin-averaged empirical spectral matrices are computed according to
eq.~(\ref{eq:binexp}) and a model of independent components is
adjusted to them by minimizing (\ref{eq:phi}).  The resulting
bin-averaged harmonic spectra of each component is then corrected by
inverting the bin-averaged version of relation~(\ref{eq:MASTER}).
This last stage also incorporates the correction of the common beam
pattern.

\subsection{Component map separation}

Ideally, component maps would be estimated by applying a Wiener filter
$W = [A^t N^{-1} A + C^{-1}]^{-1}A^t N^{-1}$ to the observations.
This solution maximizes the signal to noise ratio on each individual
component map (here noise means detector noise plus other
astrophysical component emissions). In the limiting case where noise
is small as compared to component signals, $C^{-1}$ is negligible and
the Wiener filter yields unbiased (in the sense that $W A=I$)
estimates of the maps. In poor signal to noise regimes, the signal is
attenuated to suppress noise contamination in the reconstructed maps.

In practice, the Wiener filter is applied in the frequency domain,
using estimated values $\widehat A$, $\widehat N$ and $\widehat C$ of
the parameters.  The harmonic coefficients of the estimated components
are obtained as
\begin{equation}
  \label{eq:wiener}
  \hat s (l,m) =
  [\widehat A^t \widehat N_q\inv \widehat A + \widehat C_q\inv ]\inv
  \widehat A^t \widehat N_q\inv
  x (l,m)
\end{equation}
when $\lmin(q) \leq \ell \leq  \lmax(q)$.

\section{Application \& Results}\label{sec:results}

We now present a SMICA analysis of WMAP data.
The two data sets described in section \ref{sub:maps} (map set I
corresponding to sky regions outside of Kp2 mask, and map set II
corresponding to galactic latitudes higher than 40 degrees, with the
strongest point sources masked in both cases) are used for two
independent analyses.

\subsection{Choice of parameters for SMICA}\label{sub:choiceparam}

In a preliminary analysis, we have looked for the total number of
components required by the map set~I without imposing constraints
neither on the amplitude of the estimated components nor on their
spatial power spectra.  In this totally blind analysis, we found two
significant components.
The first component, dominant in all channels, is clearly identified
as CMB.  The second, weaker by several orders of magnitude and
essentially significant in the Q band (it is detected to a lesser
extent in the V and W bands) is thought to be a mixture of residual
galactic emissions and unresolved point sources.  These last two
processes cannot be reliably separated without introducing additional
constraints because of nearly proportional mixing matrix columns (both
components dominate in the Q band and are almost negligible in the
others).

In order to differentiate between residual galactic emission and point
sources, we need to introduce some physical knowledge in the form of
constraints on the system.
Therefore, in most of the subsequent analysis, unless explicitly
stated otherwise, we adjust a model with three components, one of them
being strongly constrained to capture point source contributions as
follows.
The emission spectrum of point sources is well described, at WMAP
frequencies, by a power law with a spectral index $\beta \approx -2$
(section \ref{subsub:model}) and their spatial power spectrum is
expected to be almost flat since the effect of clustering is
negligible in practice at radio frequencies \citep{EPSS}.  
Hence, for one component, meant to be residual point source emission,
we fix the mixing parameters (the column of the mixing matrix) to
$A_{i,{\rm PS}} = (\nu_i/\nu_0)^{-2}$ in RJ temperature units, and we
constrain the harmonic power spectrum to be flat.  Only one parameter,
its amplitude, is left free to match the contribution of this
component.

In summary, we match spectral matrices with three components.  For two
of them, meant to be CMB anisotropies and residual galactic emission,
no constraint on parameters are enforced (both the mixing parameters
and the power spectra are determined exclusively from the data).  The
last component is constrained to have the emission law and spectral
shape of point sources; only its amplitude can be adjusted (but see
figure~\ref{fig:pspwsp} and the related comments).  Finally, regarding
noise, the average noise power is freely estimated in each map and
each frequency bin.

%%%%%%%%%%%%%%%%%%%%%%%%%%%%%%%%%%%%%%%%%%%
\begin{figure*}
  \begin{center}
    \includegraphics[width=2\columnwidth]{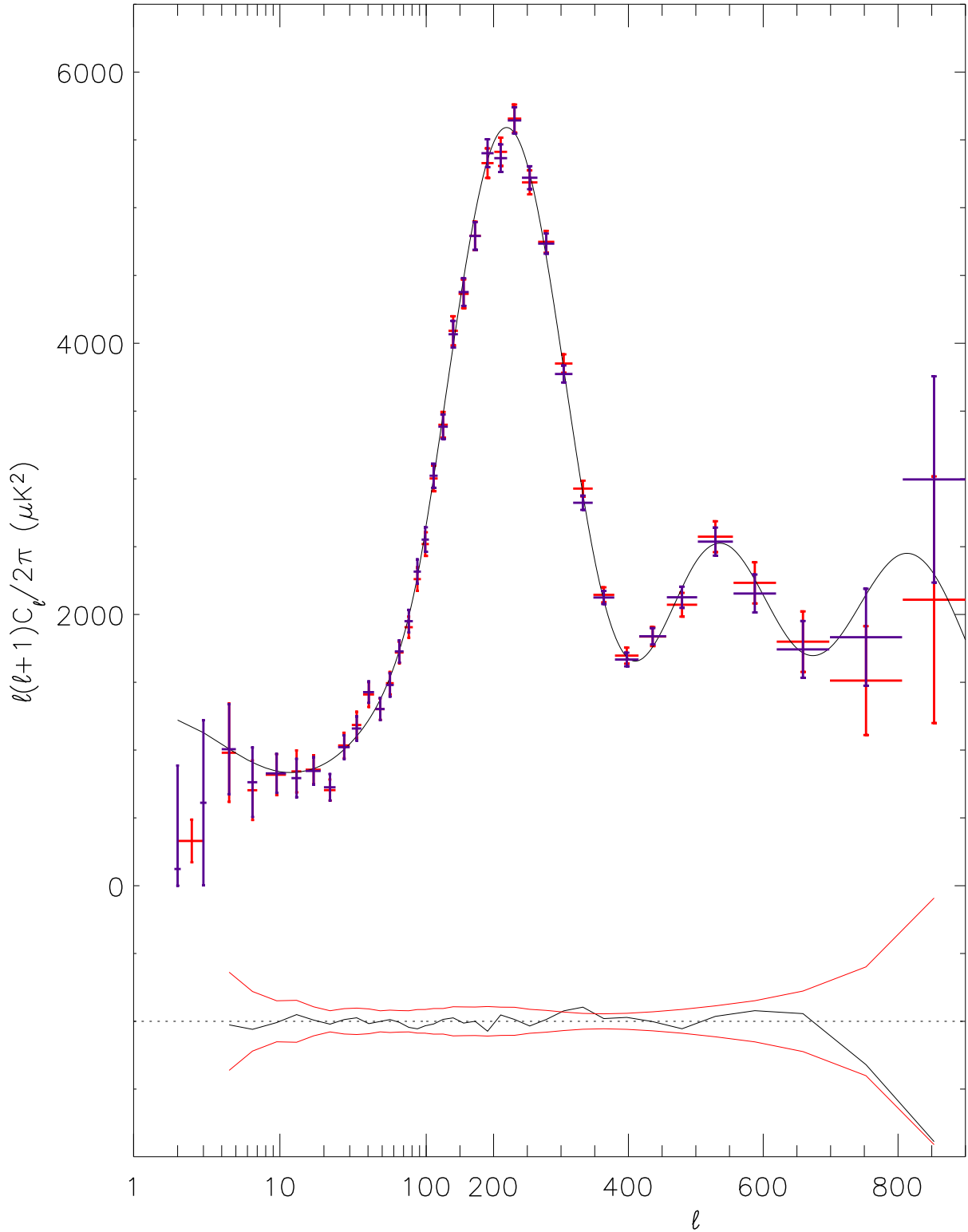}
  \end{center}
  \caption{The CMB spatial power spectrum measured with SMICA (in red)
  as compared to the published WMAP 1 year spectrum (in blue). The
  error bars from SMICA are computed from the Fisher information
  matrix using the parameters at convergence (and not the parameters
  of the model).  This explains the small error bar on the first bin
  of the SMICA estimate. We also plot the difference of the power
  spectra for better comparison. The red curves correspond to $\pm 1
  \sigma$ error bars on the power spectrum obtained with SMICA.}
  \label{fig:CMBSp}
\end{figure*}
%%%%%%%%%%%%%%%%%%%%%%%%%%%%%%%%%%%%%%%%%%%%

\subsection{CMB anisotropies}

Figure \ref{fig:CMBSp} shows the estimated CMB power spectrum after
correcting for partial coverage and beam and pixel transfer functions.
It displays a peak around $\ell = 200$ (first acoustic peak) and a
second peak around $\ell = 550$, both compatible with the measurement
announced by the WMAP team.

SMICA and WMAP team power spectrum estimations show an excellent
agreement for most multipoles. For multipoles between $\ell = 2$ and
$\ell = 290$ the difference between the two power spectra is much
smaller than error bars (except for bins centered on $\ell = 190$ and
$\ell = 210$ where the difference is of the order of the error bars),
but is larger than statistical errors after removing cosmic variance
contribution.  This can be explained (at least partially) by the small
differences in the sky coverage due to the apodisation procedure we
apply on the maps.  For larger multipoles, the two estimates show
somewhat larger differences which may be due, in part, to the two
different weighting schemes used by the two methods for $\ell > 200$.
For SMICA, no weighting of the pixels depending on the noise variance
per pixel is applied prior to the analysis.  On the contrary, the WMAP
team applies different weighting schemes for $\ell > 200$ (a weighting
proportional to 1/noise$^2$ for $\ell > 450$ and a transition
weighting for $200 < \ell < 450$).  Hence, discrepancies between the
two estimates are expected to be larger for $\ell > 200$ since the two
data sets are not quite identical. This may account for the small
observed differences.
We note in the passing that the small dip at the top of the first
acoustic peak in the WMAP team estimate has disappeared in the SMICA
estimate.

%%%%%%%%%%%%%%%%%%%%%%%%%%%%%%%%%%%%%%%%%%%%
\begin{figure}
  \begin{center}
    \includegraphics[width=\columnwidth]{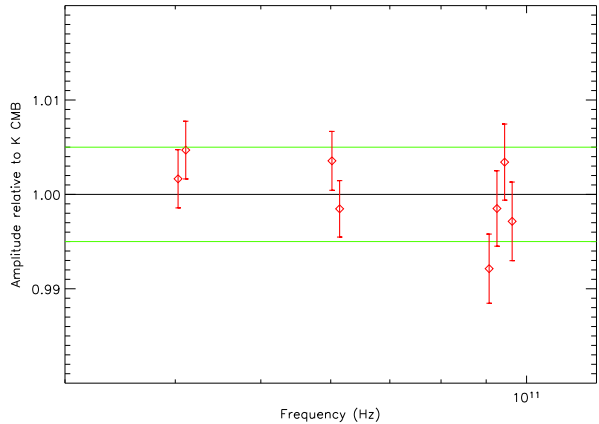}
  \end{center}
  \caption{Measured amplitude of CMB fluctuations in each detector (in
  red) relative to temperature fluctuation units as given by the
  calibration on the dipole.  The amplitude of the fluctuations are
  normalized to 1 on average. Data points at the same frequency have
  been slightly offset in abscissa for readability. Blue lines delimit
  calibration 1 sigma errors provided by the WMAP team. The errors on
  estimated mixing parameters are statistical errors computed from the
  Fisher information matrix at convergence. In order to provide an
  error bar on every CMB mixing parameter, errors are not marginalized
  on CMB power spectrum estimate (there is a degeneracy between power
  spectrum amplitude and mixing parameters normalization).}
  \label{fig:CMBmixmat}
\end{figure}
%%%%%%%%%%%%%%%%%%%%%%%%%%%%%%%%%%%%%%%%%%%%
Figure \ref{fig:CMBmixmat} shows the estimated amplitude of CMB
anisotropies for all the observation channels, as given by the CMB
column of the estimated mixing matrix $\hat A$.  We find an emission
law compatible with the expected derivative of a blackbody, to
excellent accuracy. A fit of the form $\nu^\alpha$ gives $\alpha =
-0.0067 \pm 0.0058$ compatible with 0 at about 1.15 sigma.  Refined
calibration and systematics testing, as well as the second year data,
will reduce the error bar and clarify the significance of a possible
deviation from the expected blackbody derivative.
Note that, given the error bars, \emph{the accuracy of the measurement
of the CMB emission law is limited by detector calibration uncertainty
rather than intrinsic statistical errors!}  The present measurement
shows that the emission law of the anisotropies is the same as that of
the dipole, which is itself known (under the assumption that it is
essentially due to a kinetic effect) to be the derivative with respect
to temperature of the CMB blackbody emission measured with FIRAS.

%%%%%%%%%%%%%%%%%%%%%%%%%%%%%%%%%%%%%%%%%%%%
\begin{figure*}
  \begin{center}
    \includegraphics[angle=90,width=2\columnwidth,]{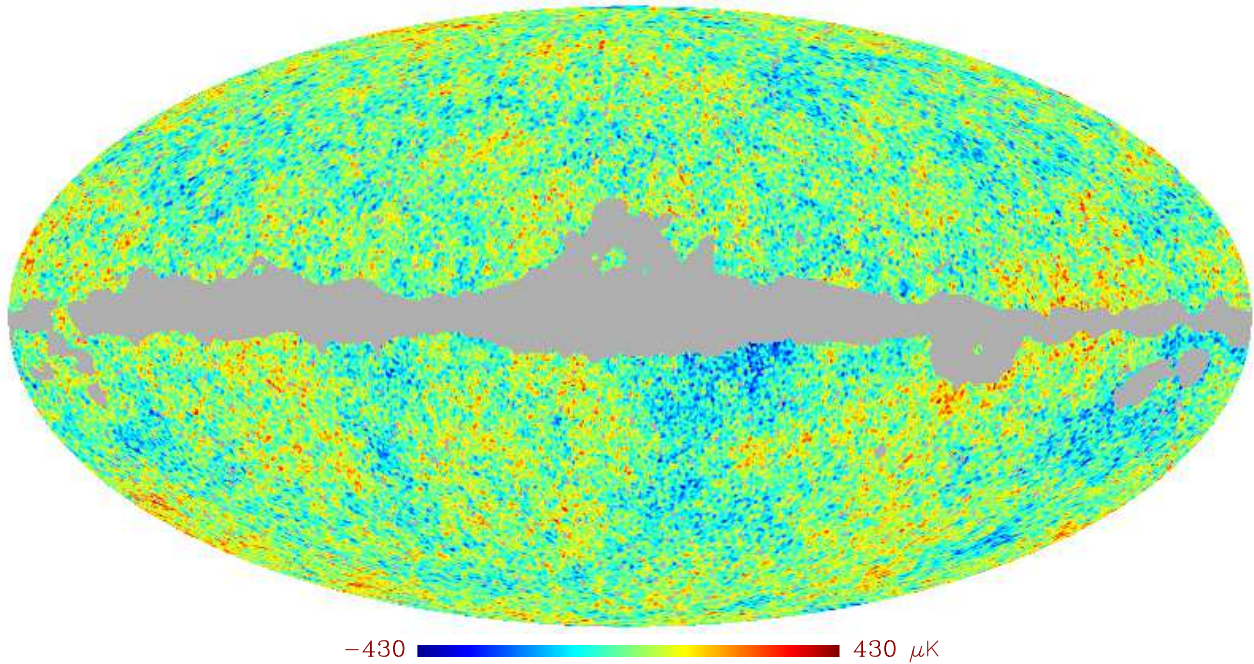}
  \end{center}
  \caption{Map of CMB fluctuations obtained after Wiener filtering of
    the data using parameters estimated with SMICA.}
  \label{fig:cmbmap}
\end{figure*}
%%%%%%%%%%%%%%%%%%%%%%%%%%%%%%%%%%%%%%%%%%%%%
Figure \ref{fig:cmbmap} displays a reconstruction of the CMB
anisotropy map by Wiener filtering (equation \ref{eq:wiener}) using
the parameters estimated with SMICA.
The now familiar CMB anisotropy pattern is clearly visible.

We have compared our recovered CMB map with other available CMB maps
from WMAP: the Internal Linear Combination (ILC) map published by the
WMAP team \citep{wmap-foreg}, the combined map used for
non-Gaussianity analysis (e.g., \cite{komatsu}) and the two maps
(\emph{cleaned} and \emph{Wiener filter}) obtained by Tegmark
\citep{tegmark03}.

\begin{figure}
  \begin{center}
    \includegraphics[width=\columnwidth]{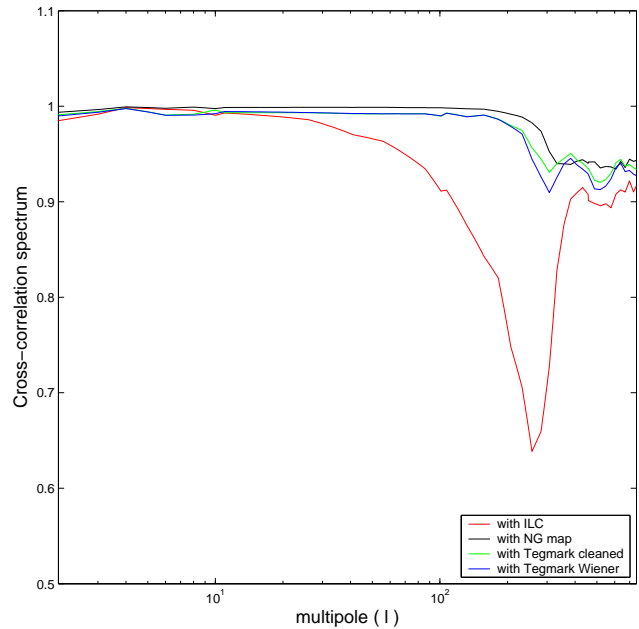}
  \end{center}
  \caption{%%
  Cross-correlation spectra between our CMB map with other available
  CMB maps from the WMAP data: the ILC map published by the WMAP team
  (Bennett et al. 2003b), the combined map used for non-Gaussianity
  studies (e.g., Komatsu et al. 2003) and the two maps (\emph{cleaned}
  and \emph{Wiener filter}) obtained by Tegmark et al. (2003).)}
  \label{fig:corrmaps}
\end{figure}

As seen on figure~\ref{fig:corrmaps}, we find very high correlation
($\approx 99\%$) with the two maps proposed by Tegmark as well as with
the map used for non-Gaussianity studies up to $\ell \approx 200$.
The correlation slightly decreases down to $\approx 95\%$ for
multipoles larger than 200 (see Figure~\ref{fig:corrmaps}).
The cross-correlation with the ILC map is also around of 99\%, but
only up to multipoles $\ell \approx 20$. For smaller scales, the
correlation with the ILC map decreases down to $\approx 60\%$ at $\ell
\approx 200$ and then it increases up to $\approx 90\%$ at the
smallest scales.  The fact that the correlation with the ILC is
smaller than with the other CMB maps is not surprising since, as
explained by the WMAP team, the ILC was developed for foreground
analyses and it is far for being the best CMB reconstruction (Bennett
et al. 2003b).
%%%%%%%%%%%%%%%%%%%%%%%%%%%%%%%%%%%%%%%%%%%%%%
%%%%%%%%%%%%%%%%%%%%%%%%%%%%%%%%%%%%%%%%%%%%%%

We have also tested possible contamination caused by unsubtracted
foreground emissions by cross-correlating our CMB reconstruction with
templates of different Galactic emissions (synchrotron, thermal dust
and free-free).  We find no significant contamination, since the
correlation level is below the dispersion of the casual correlations
between those templates and CMB Gaussian simulations ($5 - 10 \%$).

\subsection{Residual galactic emission}

Table \ref{tab:Aresidual} gives the emission law (the column of the
mixing matrix) estimated for the second component in the map set I.
%%%%%%%%%%%%%%%%%%%%%%%%%%%%%%%%%%%%%%%%%%%%
\begin{table}
  \begin{center}
    \tiny{
    \begin{tabular}{@{}lcccccccc}
      detector           & Q1&Q2&  V1 &  V2 &  W1  &  W2 & W3  & W4   
      \\\hline
      ampl. (K/K)&1&0.85& 0.04& 0.02& -0.13&-0.27&-0.22&-0.16\\\hline
      error (K/K)&0&0.08& 0.07& 0.07&  0.08& 0.09& 0.09& 0.09\\\hline
    \end{tabular}
    }
  \end{center}
  \caption{Estimated relative amplitude of the second component in
  each of the observed map in Rayleigh-Jeans (RJ) temperature units
  and associated errors. The normalization of parameters is fixed with
  respect to Q1 map. Error on parameters are correlated at the 20\%
  level.}
  \label{tab:Aresidual}
\end{table}
%%%%%%%%%%%%%%%%%%%%%%%%%%%%%%%%%%%%%%%%%%%%%
Parameters are rescaled to normalize to unity in the Q1 band (where
the component dominates) in order to fix the degeneracy between mixing
parameter normalization and power spectrum amplitude. Although much
lower than the CMB, this second component is clearly detected in the Q
band (at about 10 standard deviations). Its mixing parameters are
compatible with 0 in V-bands, and are systematically negative and
detected at about two standard deviations on average in each
individual map of the W-bands except W1.  Mixing parameters for
radiometers observing at the same frequency are compatible with a
constant value, as expected for an astrophysical component. The
amplitude of the component is at least 10 times larger in the Q band
than in the V band (in Rayleigh Jeans (RJ) temperature units), and is
larger in absolute value in the W band than in the V band. This
behavior is not {\it a priori} excluded if the component originate
from residual galactic emission (mainly synchrotron and dust
correlated emission) after galactic foreground removing operation by
the WMAP team. In that case, mixing parameters of two different
frequencies may have opposite signs depending on the sign of the error
on the parameters of the template fit.

\begin{figure*}
  \begin{center}
    \includegraphics[angle=90,width=2\columnwidth,]{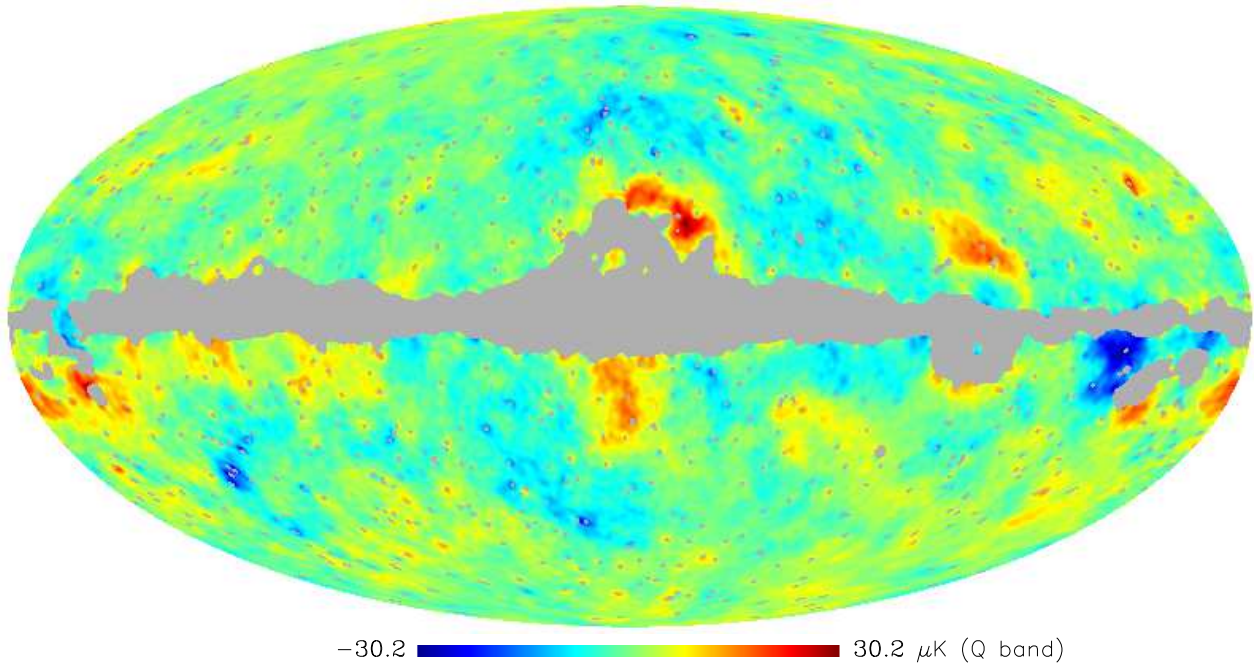}
  \end{center}
  \caption{Map of the residual component as ``seen'' in the
    Q-band, obtained by Wiener filtering using estimated parameters. }
  \label{fig:wienGal}
\end{figure*}

Figure \ref{fig:wienGal} shows the map of the residual component
obtained by Wiener filtering using estimated parameters.
%%%%%%%%%%%%%%%%%%%%%%%%%%%%%%%%%%%%%%%%%%%%%
%%%%%%%%%%%%%%%%%%%%%%%%%%%%%%%%%%%%%%%%%%%%%
We can notice a bright structure close to the galactic center at the
edges of the galactic mask which we identify as a residue from
foreground subtraction in the \emph{Ophiuchus complex}.  Other
residual structures associated to the \emph{North polar spur}, the
\emph{Gum nebula}, the \emph{Orion-Eridanus bubble} and the
\emph{Taurus} region can also be identified, suggesting that the
emission of these regions does not perfectly match the model used by
WMAP team.

We have found a significant correlation ($\approx 40\%$) between our
second component and the WMAP synchrotron emission estimation at the Q
band (using MEM, see Bennett et al. 2003b), which supports the
Galactic origin of our second component.

As can be seen from the Wiener map (and as expected for galactic
emission), the component looks very non-stationary over the sky.  As
the stationarity of components is assumed in our model, errors on
mixing parameters probably are somewhat underestimated but we do not
expect this to change our interpretation.

\begin{figure}
  \begin{center}
    \includegraphics[width=\columnwidth]{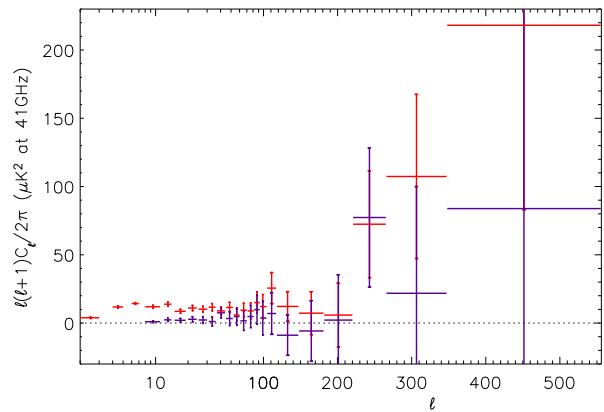}
  \end{center}
  \caption{Estimated power spectrum of the second component from
  SMICA analysis in map set I (red ink) and comparison with the power
  spectrum estimated at galactic latitudes higher than 40 degrees (map
  set II, blue ink).  Power spectra amplitudes are rescaled on the
  Q-band. No positivity constraints are enforced for parameter
  estimates, making it easy to check their compatibility with~$0$.
  Cosmic variance, naturally included in error bar computation from
  the Fisher information matrix, has been removed afterwards from the
  error estimates.  Estimated power spectra are re-binned for
  readability. Due to the lower sky coverage fraction for map set II
  (30\%), the large angular scales are poorly constrained.  We then
  restrained the analysis to multipoles $\ell > 10$ only.}
  \label{fig:galpwsp}
\end{figure}

Figure \ref{fig:galpwsp} shows the estimated power spectra of the
second component in both map sets I and II. For the estimation with
the map set II, we have fixed the mixing parameters of the residual
component, as well as the amplitude of point source power spectrum to
the results obtained in the map set I. This allows to compare the two
power spectra estimates of a component with a fixed effective emission
law.
%%%%%%%%%%%%%%%%%%%%%%%%%%%%%%%%%%%%%%%%%%%%%
%%%%%%%%%%%%%%%%%%%%%%%%%%%%%%%%%%%%%%%%%%%%%%
The estimated power spectrum of the residual component in map set~I is
measured with high significance for multipoles between $\ell =2$ and
$\ell \simeq 150$ and form a plateau (${\ell^2}c(\ell) \approx
\mathrm{const}$) with an amplitude $\approx 10-12 \, \mu {\rm K}^2$ in
the Q-band. For higher multipoles, the beam transfer function of
Q-band detectors strongly suppresses the signal, and the sensitivity
is not sufficient to allow an accurate measurement of the component.
Nevertheless, the last three bins show weak incompatibilities with 0
at 1.5-2 sigma.  Errors on those three parameters are strongly
correlated (to $\approx 50\%$ due to the anti-correlation of those
parameters with the amplitude of point sources emission), so there is
not clear evidence of detection at high spatial frequencies.

The power of the residual component for angular scales larger than 1-2
degrees is clearly reduced at galactic latitude higher than 40
degrees.  It remains marginally detectable even though it is reduced
by a factor greater than 4 for $\ell$ between about 10 and 30.

We find that the power of the remaining galactic foreground
contamination in WMAP maps (outside of Kp2 mask) is about 1\% of the
CMB anisotropy variance at large angular scales ($\ell < 100$) at
41~GHz.  
This result is in agreement with the WMAP team estimate, based on
correlation measurement using external foreground templates
\citep{wmap-foreg}.  The power of the residual component is less than
0.2 \% of the CMB variance at 61 and 94 GHz.

\subsection{Unresolved point source emissions}\label{sub:ps}

In our analysis, the third component is rigidly constrained: as
described above (sec.~\ref{sub:choiceparam}), the mixing parameters
and the shape of the power spectrum are fixed to values expected for
unresolved point sources emission, leaving only the overall amplitude
to be determined by spectral matching.
We find $C_{\mathrm{PS}}(\ell)=(9.2\pm5.0)\times 10^{-3}~\mu K^2$ at
41 GHz. This value is marginally compatible with the WMAP team
estimate (see \cite{komatsu} for a discussion for the expected
contribution of unresolved point sources to the CMB power spectrum,
for different masks and flux limits) and model predictions
\citep{Argueso}: $15.5 \pm 1.7 \times 10^{-3}\mu K^2$.

The large error bar on our estimate is due to the difficulty of
discriminating components with similar spectral characteristics.  In
this experiment, the residual galactic component and the point source
component have similar effective emission laws (mixing matrix
elements) and both components dominate in the Q-band and can not be
accurately estimated in V and W bands. So, some power at high spatial
frequency can be exchanged between the second and the third components
without modifying dramatically the likelihood of the model.

\medskip

\begin{figure}
  \begin{center}
    \includegraphics[width=\columnwidth]{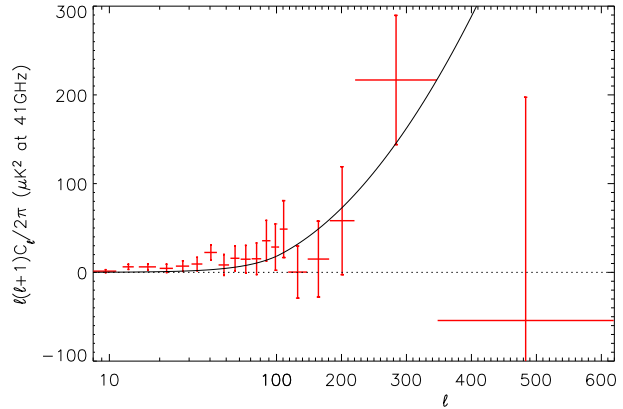}
  \end{center}
  \caption{Power spectrum of unresolved point sources emission
    estimated with SMICA assuming two components in the map set II
    (red).  Mixing parameters of point source emission have been fixed
    for the analysis. No constrains of positivity is given for
    parameter estimates allowing to check easily their compatibility
    with 0.  Cosmic variance has been removed afterwards from the
    error estimates. The black curve corresponds to the model of point
    source power spectrum $C_{\mathrm{PS}}(\ell) = const$ with an
    amplitude estimated from the semi-blind approach of $11.3 \times
    10^{-3} \mu K^2$ (see text for details).}
  \label{fig:pspwsp}
\end{figure}

We further investigate the distribution of the third component by
performing two additional spectral matches on map set~II.
Neglecting residual galactic emissions (we have seen in previous
section that residual galactic emission is clearly reduced at high
galactic latitude), we now fit a two-component model (as opposed to
previous matches, build with three components) with two different
constraint set.
In a first match, all parameters related to point sources emission law
and power spectra are fixed, except the amplitude which is fitted.  We
find the amplitude of the power spectrum to be: $C_{\rm
  PS}(\ell)=(11.3\pm3.5)\times 10^{-3}~\mu K^2$ at 41 GHz, in better
agreement with the value already reported of $15.5 \pm 1.7 \times
10^{-3}\mu K^2$.
In a second match, we relax all constraints on the shape of the power
spectrum to obtain estimates of the power spectrum in every multipole
bin.
The resulting power spectrum of unresolved point sources is displayed
in figure~\ref{fig:pspwsp}.
%%%%%%%%%%%%%%%%%%%%%%%%%%%%%%%%%%%%%%%%%%
%%%%%%%%%%%%%%%%%%%%%%%%%%%%%%%%%%%%%%%%%%%%
Parameters are clearly incompatible with 0 for most of the multipoles
(assuming {\it a priori} a smooth power spectrum).  A good
compatibility with the model $C_{\mathrm{PS}}(\ell) = 11.3 \times
10^{-3} \mu K^2$ (41 GHz) is observed. If we re-adjust the model of
the power spectrum to the estimated parameters using a simple $\chi^2$
we obtain a similar value for the amplitude. Nevertheless, a weak
excess of power can be seen for the lowest multipoles ($\ell < 50$)
possibly due to a remaining galactic contamination at high galactic
latitudes.

\subsection{Goodness of fit}\label{sub:FitRes}

In this section, we briefly examine the fit of the best models to the
data across frequency bins by plotting the weighted mismatch $n_q
D(\widehat R_q, R_q(\hat\theta))$ against the bin index $q$.  
If the model of independent stationary component holds, the expected
value of this measure of mismatch is given, for $n_q$ large enough, by
eq.~(\ref{eq:Kfit}).  In our plots, however, we use Monte-Carlo
simulations to have an estimate of the distribution of the mismatch
valid even in the non asymptotic regime.
We report goodness of fit of adjusted models in two cases: a
one-component model and a three-component model.

\begin{figure}
  \begin{center}
    \includegraphics[width=\columnwidth]{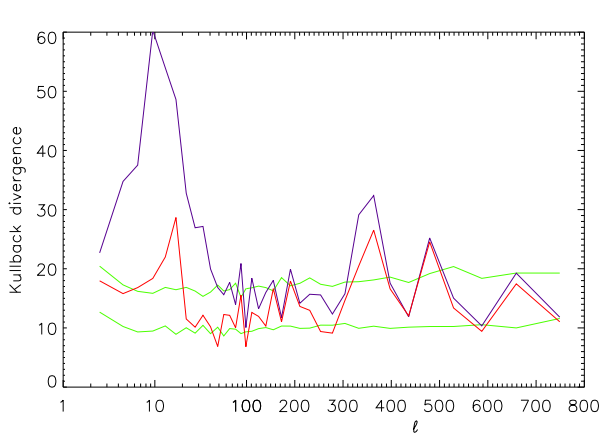}
  \end{center}
  \caption{Spectral mismatch for map set I, with three components  
    (red) and 1 component (blue). The green curves are the boundaries
    of the 68\% goodness of fit interval estimated using simulations.
    For three components, the fit is very satisfactory for most of the
    spatial frequencies (see text).}
    \label{fig:CMBSpKdiv}
\end{figure}

Figure~\ref{fig:CMBSpKdiv} shows that the one-component model is
clearly incompatible with the data for $\ell < 60$.  It also shows
that the fit obtained with three components is very satisfactory for
most of the multipoles although small discrepancies remain, in
particular for $\ell$ between 10 and 25.  We interpret this as a clear
indication that residual galactic emission are present in the data at
low spatial frequency.

%%%%%%%%%%%%%%%%%%%%%%%%%%%%%%%%%%%%%%%%%%%%%%
%%%%%%%%%%%%%%%%%%%%%%%%%%%%%%%%%%%%%%%%%%%%%

\begin{figure}
  \begin{center}
    \includegraphics[width=\columnwidth]{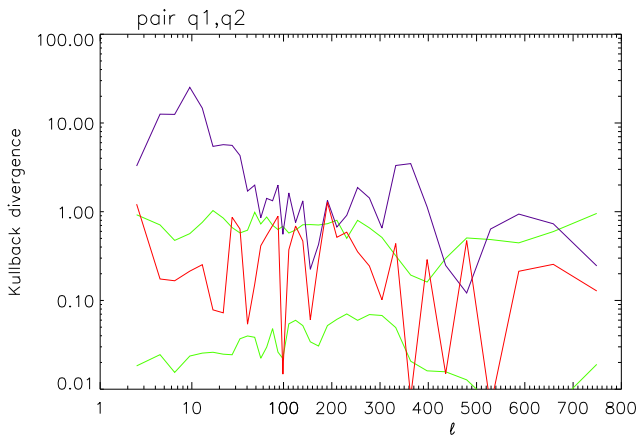}
  \end{center}
  \caption{Spectral mismatch for the detector pair (Q1,Q2) only, with  
    three components (red) and one component (blue). The green curves
    are the boundaries of the 68\% interval estimated using
    simulations. Note the logarithmic scale and thus the very
    significant reduction of the mismatch when three components are
    considered.}
    \label{fig:KdivQ1Q2}
\end{figure}

We can investigate the origin of the remaining discrepancies by
looking at the spectral mismatch between \emph{pairs} of detectors.
Figures~\ref{fig:KdivQ1Q2} shows the goodness of fit between the
individual pair of detectors (Q1,Q2), assuming either one or three
components.  The misfit for the one-component model is now even more
obvious (as compared to the global mismatch involving all detectors),
as expected since the Q band is where the galactic residual dominates.
%%%%%%%%%%%%%%%%%%%%%%%%%%%%%%%%%%%%%%%%%%%%%
%%%%%%%%%%%%%%%%%%%%%%%%%%%%%%%%%%%%%%%%%%%%%%

As the second component is detected in the W channel as well, we also
looked at the fit of one-component and three-component models for all
W detectors together and for the set of W2, W3 and W4 (discarding W1).
For both figures, the goodness of fit is significantly better for
three components than for a unique component for low multipoles.

\begin{figure}
  \begin{center}
    \includegraphics[width=\columnwidth]{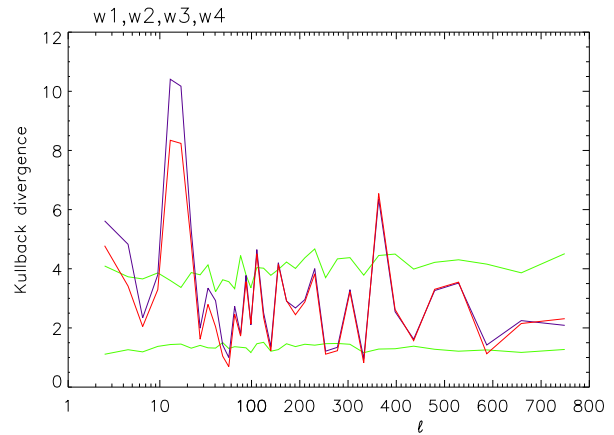}
    \\
    \includegraphics[width=\columnwidth]{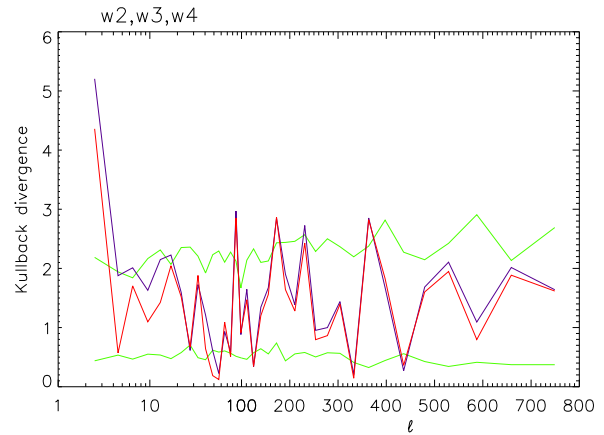}
  \end{center}
  \caption{%%
  Spectral mismatch with three components (red) and one component
  (blue).  Top: mismatch for all four W band detectors; bottom: for
  three last W band detectors (W1 excluded).}
  \label{fig:Kdivwwww}
\end{figure}

A significant excess remains, which is probably due to fine departure
of the residual galactic emission from the single template assumption.
Figure~\ref{fig:Kdivwwww}, however, rather point to remaining
systematics in the W1 channels since the mismatch is significantly
reduced when excluding the W1 detector.

This incompatibility, however, is small enough so that the CMB power
spectrum is not affected because the errors for the lowest multipoles
are dominated by cosmic variance.

%%%%%%%%%%%%%%%%%%%%%%%%%%%%%%%%%%%%%%%%%%%%%%%%%%%%%%%%%%%%%%%%
%%%%%%%%%%%%%%%%%%%%%%%%%%%%%%%%%%%%%%%%%%%%%%%%%%%%%%%%%%%%%%%%

\section{Conclusions}\label{sec:conclusion}

We have performed a blind multi-component analysis of the first year
WMAP data.  We used the eight foreground-cleaned high frequency maps
(Q1 to W4) provided by the WMAP collaboration.
Our analysis uses SMICA, a maximum likelihood spectral matching
method.  Three significant astrophysical components
\begin{itemize}
  \item the CMB anisotropies
  \item a residual galactic emission
  \item unresolved point source emissions
\end{itemize}
are jointly characterized in the foreground cleaned maps after masking
the strongest point sources and the galactic plane using Kp2 mask.
No significant thermal SZ emission is found with the present study, in
agreement with the WMAP team \citep{wmap-foreg} and \cite{SZ-WMAP}.

Blind analysis allows us to estimate the power spectrum of CMB
anisotropies as well as their amplitude across frequency channels
confirming their cosmological nature. 
Our power spectrum estimate is in excellent agreement with the WMAP
team estimate. We show that the measured emission law of CMB
anisotropies at WMAP frequencies is compatible with the derivative of
a blackbody, as expected for temperature fluctuations. The statistical
errors on parameters related to the amplitude of the anisotropies are
about 0.3\%, and are smaller than calibration errors (0.5\%) on dipole
modulation. Conversely, if CMB anisotropies are assumed to be pure
temperature fluctuations, then the estimation of CMB amplitude across
channels provides a relative calibration across frequency bands at a
better precision than dipole calibration.

The second estimated component, corresponding to a weak residual
galactic emission, is mainly concentrated in Q-band maps outside of
the Kp2 mask.  We believe that this component results mainly from
spatial variations of the difference between Haslam map (used as a
synchrotron template for the subtraction) and synchrotron emission at
WMAP frequencies.  This component is weak compared to CMB
anisotropies.  The estimated power spectrum is about
$\ell(\ell+1)C(\ell)/2\pi \approx 10-12\mu K^2$ for $\ell < 100$ in Q
band and less than $2\mu K^2$ in V and W band. Those estimates are
compatible with WMAP team estimates of foreground contamination.
Finally, much of the power from this residual large scale component
disappears for galactic latitude higher than 40 degrees.

The third component corresponds to residual point sources emission.
By fixing the parameters related to the emission law of point sources
(we assume $(\nu/\nu_0)^{-2}$) and to the power spectrum (flat power
spectrum), we estimate the amplitude of unresolved point source power
spectrum. We find at high galactic latitude: $C(\ell) = (11.3\pm3.7)
\times 10^{-3}\mu K^2$ at 41~GHz compatible with the WMAP team
estimation.  We also provide an estimate of the power spectrum of
point sources at high galactic latitude.

The goodness of fit of the 3-component model is excellent, except for
a very small discrepancy around $\ell\approx 20$ which is likely to be
attributed to a systematic in W1.  This inconsistency has no impact on
the CMB power spectrum estimate (because errors are dominated by
cosmic variance), but could affect the estimates of weak components in
data.

\section{Acknowledgments}

This work was supported by the Canadian Natural Sciences and
Engineering Research Council and by the French ministry of research.
GP would like to thank Mark Halpern for useful discussions.  We
acknowledge the use of the Legacy Archive for Microwave Background
data analysis (LAMBDA). Some results of this paper have been derived
using the HEALPix \citep{healpix} package.  We acknowledge the use of
the software package CMBFAST (http://www.cmbfast.org) developed by U.
Seljak and M. Zaldarriaga.

\label{lastpage}

\end{document}